# Quantum Memory and Quantum Cloning in Atomic Frequency Comb

Mingzhen Tian and Devin Vega

Department of Physics and Astronomy, Quantum Materials Center

George Mason University, Fairfax, VA 22030

**Abstract**: Atomic frequency comb (AFC) made of an ensemble of atoms with a periodic optical resonance was originally proposed as a viable approach for quantum memory for photons. In this paper, we examine the quantum cloning capacity of an AFC in terms of the spectral distribution of the atomic populations in the energy levels associated by the optical transition. Expressions are derived for the memory readout efficiency, signal to noise ratio, and fidelity for an input at the single photon level. When applied to a square toothed AFC, our analysis shows that there is a region where amplification from the excited state results in greater than unit recall efficiency, while maintaining a fidelity greater than the classical limit.

## I. Introduction

Optical quantum memory at the light-matter interface has been widely recognized as a critical technology for quantum information processing and communication [1-5]. Rare-earth ensembles trapped in inorganic crystals have been extensively studied in recent years as one of the promising solid-state systems to implement quantum memory with goals of achieving high efficiency and fidelity, as well as long storage time, multimode capacity, high time-bandwidth product, and on-demand retrieval [6]. Many protocols have been developed, including CRIB [7-9], GEM [10], AFC [11], and ROSE [12] based on variations of the photon echo process in atomic ensembles with inhomogeneously broadened optical resonances. Demonstrations have shown the desired memory properties listed above in separate systems. However, the demonstration of all these properties in one system has not yet been successful. All photon-echo based approaches so far have been designed on a common feature—the inherently high fidelity of the retrieved photon compared with the input photon. This is achieved by preventing atoms from populating the exited state at the time of echo emission, thus avoiding noise due to spontaneous emission. Although each of these protocols has its own challenges to overcome, major efforts have been focused on improving memory efficiency utilizing atoms in the ground state only. On the other hand, it is well known that an excited state atomic population can boost echo efficiency in photon-echo based classical signal processing [13-16]. Unfortunately, a regular two-pulse or three pulse echo process with significant excited state population cannot be used as a quantum memory because the spontaneous emission causes degradation of the fidelity below the classical limit at the single photon level [17, 18]. While existing memory protocols all weigh on eliminating excited state population to avoid spontaneous emission, the effect of the excited state population has not been studied. Such effect on memory efficiency and fidelity is especially interesting in AFC memory since the preparation of an AFC involves pumping unwanted atoms from the ground to excited states, eventually to a long-lived metastable state. Atoms in the excited state may be helpful in improving memory efficiency, but they should also be considered to account for potentially undesirable residual undecayed population.

An inverted medium with more excited population than the ground state can be an amplifier: An input optical signal is amplified through stimulated emission and noise is added to the output due to spontaneous emission. In the quantum regime, a photon in an arbitrary state cannot be duplicated without degradation of its fidelity. This is fundamentally governed by the no-cloning theorem [19, 20]. Nevertheless, it is allowed that an arbitrary quantum state may be cloned with certainty and imperfect fidelity, or with less than unit probability and perfect fidelity. Both types of quantum cloning machines find applications in quantum information processing and communication [21, 22]. A trivial quantum cloning scheme is to measure the quantum state in an arbitrary eigenbasis and make copies according to the measurement result. This sets the lower bound (or the classical limit) for the fidelity of a useful quantum cloning machine [22-24]. The measurement–based cloning can also be regarded a trivial case of quantum memory with arbitrarily long memory time. Therefore the classical limit of the fidelity should also apply to the quantum memory. It has been proven that quantum cloning can have a fidelity better than the classical limit. The upper bound of the fidelity for a universal cloning machine that produces identical output qubits from a pure input state depends on the number of the input and output qubits [24,25]. An inverted medium is one of the optimal cloning machines whose fidelity can approach this upper limit [26-29]. An AFC made of both ground state and excited populations can be treated as a quantum cloning machine and the regenerated echo photons as clones of the input with a time delay. The input photon itself may be amplified without the delay as in a regular amplifier as well. However, we will focus on how the efficiency and fidelity of the echo depend on the atomic population in the excited state for both absorptive and amplifying AFC. A relation between fidelity and efficiency will be established. The AFC will be analyzed in terms of balancing the efficiency and fidelity of quantum memory and compared with an optimal quantum cloning machine.

## II. An AFC in both ground and excited states

We will consider an AFC consisting of an ensemble of atoms with the ground and excited states, $|g\rangle$ and $|e\rangle$, as shown in Fig. 1a. The $|e\rangle \leftrightarrow |g\rangle$ transition of the ensemble is inhomogeneously broadened with the resonance of an atom denoted by its frequency detuning $\Delta$ with respective to that of an incoming photon wave packet. The AFC is formed by both ground and excited atomic populations. The probability that an atom is in either of the states is denoted by $p_g(\Delta)$ and $p_e(\Delta)$, respectively, as functions of frequency. The AFC as a mixed state of the ensemble can be defined by a density operator:

$$\rho(\Delta) = p_g(\Delta)|g\rangle\langle g| + p_e(\Delta)|e\rangle\langle e| \qquad (1)$$

The comb shape of the spectral distribution of atoms in the ground state can be described by

$$p_g(\Delta) = b + (1-b)\xi(\Delta) * d(\Delta) \qquad (2)$$

where the Dirac comb function $d(\Delta) = \sum_{n=-\infty}^{\infty} \delta(\Delta - n\Gamma)$ sets the AFC's periodic structure defined by a uniform frequency spacing $\Gamma$ between the comb teeth. The tooth shape is defined by the normalized single tooth function $\xi(\Delta)$ and a flat background $b$ on the spectrum.

In an untreated atomic ensemble, all atoms are in the ground state with a flat spectrum ($p_g(\Delta) = b = 1$). The ground state AFC in equation (2) is usually created by optically pumping unwanted atoms to the excited state according to the atomic resonance, which then relax to a long-lived metastable state $|m\rangle$ as shown in Fig. 1a. It is convenient to assume a uniform spectral distribution of the ensemble before pumping and normalize the comb function $p_g(\Delta)$ to its peaks, where the atoms are not affected.

In a two-level system, or three level system with an undecayed excited state, the population is conserved in the ground and excited states, ie. $p_g(\Delta) + p_e(\Delta) = 1$. This leads to a period spectral distribution of atoms in the excited state $p_e(\Delta) = 1 - p_g(\Delta)$. Examples of $p_g(\Delta)$ and $p_e(\Delta)$ are sketched in Fig.1b. The ground state population takes the comb shape while the excited state shows a reversed comb shape. We will call both AFCs.

In a three-level system, the atoms in the excited state are expected to decay to the metastable state. After a portion $0 \leq 1-\chi \leq 1$ of the excited state population has decayed to the metastable state, the AFC in the excited state becomes

$$p_e(\Delta) = \chi[1 - p_g(\Delta)] = \chi(1-b)[1 - \xi(\Delta) * d(\Delta)] \tag{3}$$

The AFCs defined in equations (2) and (3) include the case of the original proposal in ref [11] where $\chi = 0$ so that all relevant atoms are in the narrow teeth in the ground state. The other extreme is that $\chi = 1$ where all of the atoms in the ensemble contribute to the AFC either in ground or excited state. More realistic cases fall in between.

### III. Photon Echo, Spontaneous Emission, and Absorption

After the AFC is set by the optical pumping and/or population relaxation, an input pulse of a few photons at $t = 0$ interacts with the AFC in the mixed state described by equation (1). As the result, a photon echo pulse is generated at $t = 2\pi/\Gamma$ containing the retrieved photons. This is similar to the three-pulse photon echo process described in [18] where an input photon and a $\pi/2$ classical pulse create a mixed state in the ensemble and another $\pi/2$ pulse interacts with the mixed state to generate an echo. However, there is a major difference in the conventional three-pulse echo compared to the echo in AFC: The input photons form one of the pulses that creates the atomic ensemble's mixed state while the AFC is created in a separate process, independent of the input photon. Another important difference is that the pulse responsible for regenerating the echo is a strong classical pulse in ref [18] compared to the few input photons in the AFC. Nonetheless, we can follow a similar theoretical treatment to analyze the process, including absorption of the input photon, generation of the echo, and spontaneous emission. In fact, the process in which we are interested in this paper is simpler than that in ref [18]: We only need to focus on the regeneration of the photon echo using a preset mixed state atomic ensemble described by equations (1) through (3). The only input pulse is the wave packet of the few incoming photons near the $|e\rangle \leftrightarrow |g\rangle$ resonance.

The interaction Hamiltonian between atomic transition $|e\rangle \leftrightarrow |g\rangle$ of dipole moment $\mu$ and the electric field $E\cos\omega t$ of an input pulse is

$$H = \frac{\hbar}{2}\begin{pmatrix} 0 & -\mu E/\hbar \\ -\mu E/\hbar & 0 \end{pmatrix} \tag{4}$$

The corresponding evolution operator is $U = \exp(-iH\tau/\hbar)$ for a pulse duration $\tau$. Assuming that the input pulse to be stored is brief and weak so that the pulse area satisfies $\theta = \mu E \tau / \hbar \ll 1$, the evolution operator becomes a rotation by a small angle $\theta = 2\varepsilon$:

$$U = \begin{pmatrix} 1 & i\varepsilon \\ i\varepsilon & 1 \end{pmatrix} \tag{5}$$

Without an input field, the atomic ensemble evolves due to the relative detuning during a time span $t$ according to a free rotation operator,

$$U_t = \begin{pmatrix} 1 & 0 \\ 0 & e^{-i\Delta t} \end{pmatrix} \tag{6}$$

And the density matrix of the ensemble after interacting with the input pulse followed by a free rotation becomes

$$\rho(\Delta,t) = U_t U \rho(\Delta) U^+ U_t^+ = \begin{pmatrix} p_g + \varepsilon^2 p_e & -i\varepsilon e^{i\Delta t}(p_g - p_e) \\ i\varepsilon e^{-i\Delta t}(p_g - p_e) & \varepsilon^2 p_g + p_e \end{pmatrix} \tag{7}$$

Assuming a uniform spectrum of the input pulse much wider than the comb spacing, the corresponding macroscopic polarization can be calculated as

$$\Pi(t) = \mu N \int \langle e|\rho|g\rangle d\Delta = i\mu N\varepsilon \int e^{-i\Delta t}(p_g - p_e) d\Delta \tag{8}$$

where $N$ denotes the spectral density of atoms in the ensemble. Both $p_g$ and $p_e$ are periodic functions of frequency that repeat at the comb spacing $\Gamma$ as described in equations (2) and (3). The induced polarization in equation (8) consists of a sequence of pulses centered periodically at $2n\pi/\Gamma$ for $n = 0,1,2...$. The assumption that the spectrum of the incoming wave packet covers several teeth of the frequency comb also implies that the time separation $2\pi/\Gamma$ between the pulses is longer than the single pulse duration. Therefore, the discrete peak value of the pulses in equation (8) can be calculated by Fourier transform of the AFC functions.

$$\Pi_n = i\mu N \varepsilon (\tilde{p}_{gn} - \tilde{p}_{en}) \tag{9}$$

where $\Pi_n = \Pi(t = 2n\pi/\Gamma)$. The Fourier components of the AFCs are defined as

$$\tilde{p}_{gn} = \int p_g(\Delta) e^{-i2n\pi\Delta/\Gamma} d\Delta = m[b\Gamma\delta(n) + (1-b)a_n\gamma] \tag{10}$$

$$\tilde{p}_{en} = \int p_e(\Delta) e^{-i2n\pi\Delta/\Gamma} d\Delta = \chi[m\Gamma\delta(n) - \tilde{p}_{gn}] \tag{11}$$

where the Fourier transform of single tooth function is denoted by $a_n\gamma = \int_{-\infty}^{\infty} \xi(\Delta) e^{-i2\pi n\Delta/\Gamma} d\Delta$. The coefficient $a_n$ depends on the single tooth function with $\gamma$ as its full width at half maximum (FWHM). The total number of teeth covered by the input photon spectrum is denoted by $m$.

The induced polarization in equation (8) includes all impulsive responses from the atomic ensemble when interacting with the input pulse. The induced electric field is proportional to $i\Pi(t)$, which is also a sequence of pulses centered periodically at $t = 2\pi n/\Gamma$ with amplitudes proportional to equation (9). The overall output field is the sum of the input and the impulsive feedback from the atomic ensemble. The feedback pulse centered at $t = 0$ with peak amplitude proportional to $(\tilde{p}_{e0} - \tilde{p}_{g0})$ is in the same spatiotemporal mode as the input and has the same timing as well. The output field at $t = 0$ (the feedback added to the input) can be regarded as the transmitted input that can be attenuated or amplified depending on the sign of $(\tilde{p}_{e0} - \tilde{p}_{g0})$. According to equations (10) and (11), $N\tilde{p}_{e0}$ and $N\tilde{p}_{g0}$ are the total numbers of atoms in the excited and ground states, respectively. The ground state population absorbs and reduces the field while the excited state population adds to the field through stimulated emission and causes noise due to spontaneous emission. This process can be treated as universal symmetric quantum cloning in an inverted medium [29]. However, we are interested in the gain and fidelity of the regenerated photons in the pulse centered at $t = 2\pi/\Gamma$.

The peak intensity of the photon echo emitted at $t = 2\pi/\Gamma$ can be written as

$$I_1 = i_0 N^2 \varepsilon^2 (\tilde{p}_{e1} - \tilde{p}_{g1})^2 \qquad (12)$$

where $i_0$ is the intensity absorbed or emitted by a single atom in the same spatiotemporal mode as the input photon. The photons in the echo pulse have the same spatiotemporal mode and quantum state, such as polarization, of the input except for a time delay. Both ground and excited state populations contribute to the number of photons in the echo output.

Besides the pulses generated from the atomic ensemble included in equation (8), the excited state population $N\tilde{p}_{e0}$ also causes spontaneous emission, adding more photons to the output. The intensity from the spontaneous emission into the same spatiotemporal mode as the echo at $t = 2\pi/\Gamma$ can be expressed as

$$I_s = i_0 N\tilde{p}_{e0} \qquad (13)$$

which is named noise in the sense that it always exists regardless of the input. The noise is the same for all possible polarizations, or qubit states, in general. With equations (12) and (13), we will further analyze the signal-to-noise ratio and fidelity of the photons in the echo pulse.

It is well-known that the storage efficiency of a ground state AFC without the excited population is maximized only if the input pulse is completely absorbed [11]. In order to study the impact from atoms in the excited state, we always compare the echo intensity with the input intensity absorbed by the ground state population:

$$I_{abs} = i_0 N^2 \varepsilon^2 \tilde{p}_{g0}^{\,2} \tag{14}$$

which is independent of the excited population. This is also treated as the input to the quantum cloning machine that generates the clones in the echo.

### IV. Memory Readout Efficiency, Signal-to-Noise Ratio, and Fidelity

In a quantum memory, the memory efficiency can be treated as the product of the storage and readout efficiencies. The storage efficiency can be defined by the ratio between the numbers of absorbed and input photons while the readout efficiency is the ratio between the numbers of output and absorbed photons. The readout efficiency of an AFC can be calculated as the ratio between the echo intensity in equation (12) and the absorbed intensity in equation (14), which relates to the AFC's Fourier components in equations (10) and (11) as

$$\eta = \frac{I_1}{I_{abs}} = \left(\frac{\tilde{p}_{e1} - \tilde{p}_{g1}}{\tilde{p}_{g0}}\right)^2 = \left(\frac{(1+\chi)(1-b)a_1}{b\Gamma/\gamma + (1-b)a_0}\right)^2 \tag{15}$$

In this way, the number of photons retrieved in the form of the echo is always compared to the same number of absorbed photons for a given AFC structure defined in equation (2). This is also consistent with the efficiency definition for the case with an empty excited state. The efficiency does not depend on the input. The fraction $0 \le \chi \le 1$ signifies the ratio between the number of atoms in the excited state and those removed from the ground state. Equation (15) shows the efficiency proportional to $(1+\chi)^2$. When the fraction increases from $\chi = 0$ (all excited atoms decayed to the metastable state $|m\rangle$) to $\chi = 1$ (all excited atoms in $|e\rangle$), the efficiency is enhanced by a factor of four regardless the AFC's actual shape.

The signal-to-noise ratio (SNR) in the echo output is defined as

$$SNR = \frac{I_1}{I_s} = \frac{N\varepsilon^2(\tilde{p}_{e1} - \tilde{p}_{g1})^2}{\tilde{p}_{e0}} \tag{16}$$

This shows that high input supports high SNR. An input pulse at the single photon level results in the lowest SNR. We will analyze the case where the single input photon is fully absorbed by the ground state population, ie. $N\tilde{p}_{g0}\varepsilon^2 = 1$. The SNR becomes

$$SNR = \frac{(\tilde{p}_{e1} - \tilde{p}_{g1})^2}{\tilde{p}_{g0}\tilde{p}_{e0}} = \frac{\tilde{p}_{g0}}{\tilde{p}_{e0}}\eta = \frac{(1+\chi)^2(1-b)a_1^2}{\chi[b\Gamma/\gamma + (1-b)a_0](\Gamma/\gamma - a_0)} \tag{17}$$

Both readout efficiency and SNR depend on Fourier coefficients $a_0$ and $a_1$ of the single tooth function, background level $b$, and finesse $\Gamma/\gamma$ of the AFC in the ground state, as well as on the fraction of population remaining in the excited state $\chi$. The fidelity of the quantum state of an output photon can be calculated from the SNR by [18, 29]

$$F = (SNR+1)/(SNR+2) \tag{18}$$

In an ideal quantum memory where any output photon carries the same input state, $SNR$ approaches infinity and $F=1$. This can only be achieved when the excited state is unoccupied ($\chi=0$). However, this does not make a perfect quantum cloning machine since the ratio of output and input cannot be more than unity. Equation (17) indicates that SNR is proportion to $(1+\chi)^2/\chi$, which reduces monotonically with $\chi$ and minimizes at $\chi=1$. A quantum memory will lose its meaning if its fidelity is lower than the maximum threshold that can be achieved by a classical memory. A classical storage of an arbitrary qubit is optimal at $SNR=1$, which sets a lower bound of fidelity $F_{cl}=2/3$ for a meaningful quantum memory. This lower bound is also consistent with the fidelity of an optimal measurement-based quantum cloning [22-24].

The background $b$ in the ground state AFC's spectrum causes a drop in both readout efficiency and SNR. At the extreme: $b=1$, the periodic structure of the AFC disappears, no photon echo is generated, and efficiency vanishes. The best case is an AFC without background ($b=0$) regardless of the tooth shape and finesse, in which the upper limits of readout efficiency and SNR are, respectively,

$$\eta = (1+\chi)^2 \left(\frac{a_1}{a_0}\right)^2 \tag{19}$$

$$SNR = \eta \frac{a_0}{\chi(\Gamma/\gamma - a_0)} \tag{20}$$

In general, Fourier transform of any single tooth function will results in $a_1 \leq a_0$ provided the comb spacing is at least twice the tooth width ($\Gamma/\gamma \geq 2$). Optimized tooth shape and finesse can make $a_1 \approx a_0$. Under such a condition, the readout efficiency can vary between 100% and 400% depending on the fraction of excited state population $\chi$. It should be noted that 100% readout efficiency at $\chi=0$ corresponds to the upper limits of the memory efficiency in the original proposal in ref [11], which are 54% and 100%, respectively, for forward and backward retrieval. The difference is caused by the propagation effect. In ref [11], the atoms in an AFC spread out in space along the path of the pulse propagation. An input photon has a higher chance to interact with the atoms at the front end than those at the back end of the ensemble due to the absorption. In a forward retrieval process, an echo photon generated at the front end has higher chance to be reabsorbed by the ensemble than one generated at the back end. As result, the memory efficiency can only be optimized to 54%. The analysis in this paper does not consider the propagation effect, where an input photon interacts with all atoms in the ensemble with an equal probability. The region $\eta > 1$ corresponds to possible quantum cloning provided the fidelity is above the classical limit.

The SNR deteriorates while the excited state population helps to increase the efficiency. The SNR reduces to $4a_0/(\Gamma/\gamma - a_0)$ with an excited state fully populated between the comb teeth as shown in Fig.1b. However, it is still possible to choose a tooth function and finesse to make the $SNR \geq 1$ so that the fidelity is still above the classical limit. We will discuss the case of an AFC of square-shaped teeth in the following section.

## V. AFC with Square teeth

A square function is one of the tooth shapes for optimal memory efficiency in a ground-state-only AFC. The single tooth is defined by the function:

$$\xi(\Delta) = \begin{cases} 1 & |\Delta| \leq \gamma/2 \\ 0 & |\Delta| > \gamma/2 \end{cases} \tag{21}$$

The relevant Fourier coefficients are calculated:

$$a_0 = 1 \tag{22}$$

$$a_1 = \mathrm{sinc}(\pi\gamma/\Gamma) \tag{23}$$

where $\mathrm{sinc}(x) = \sin x / x$.

The readout efficiency and SNR become, respectively,

$$\eta = [\frac{(1+\chi)(1-b)}{b\Gamma/\gamma + 1 - b} \mathrm{sinc}(\pi\frac{\gamma}{\Gamma})]^2 \tag{24}$$

$$SNR = \frac{(1+\chi)^2(1-b)}{\chi(b\Gamma/\gamma + 1 - b)(\Gamma/\gamma - 1)} \mathrm{sinc}^2(\pi\frac{\gamma}{\Gamma}) \tag{25}$$

The efficiency is plotted in Fig.2 at various background levels and comb finesse for a fully decayed excited state ($\chi \to 0$). The efficiency strongly depends on the background and finesse. The efficiency approaches 100% provided both zero background ($b = 0$) and high finesse (

$\Gamma/\gamma \gg 1$) are satisfied. According to equation (25), it is also confirmed that SNR approaches infinity and $F = 1$ regardless of the finesse and a background $0 \leq b < 1$.

With an undecayed excited state ($\chi = 1$), both efficiency and fidelity depend on the background and finesse. The efficiency plot has the same shape as Fig. 2 with the efficiency axis scaled up by a factor of four. The fidelity plotted in Fig. 3 decreases with the increase of both background and finesse. At zero background, the efficiency increases from 162% at $\Gamma/\gamma = 2$ to 400% at $\Gamma/\gamma \to \infty$ while the fidelity degrades from 0.724 at $\Gamma/\gamma = 2$ to the classical limit at around $\Gamma/\gamma = 4$, and eventually to 0.5 at $\Gamma/\gamma \to \infty$.

As mentioned above, perfect fidelity is only possible with an empty excited state. However, it is still possible for the fidelity to be better than the classical limit while the readout efficiency is larger than the optimal efficiency from a ground state AFC ($\eta \geq 100\%$). Fig. 4 shows the fidelity vs. efficiency for zero background at different fractions of the excited state population and the comb finesse. The top curve represents the case of an empty excited state where the fidelity is constant $F = 1$ and the efficiency $\eta \leq 100\%$, depending on the finesse. All curves are monotonic. On any given curve with fixed $\chi$, high finesse always results in high efficiency and low fidelity. The shaded area indicates the region where both $F > 2/3$ and $\eta \geq 100\%$ are satisfied. The plot shows that in the full range of $0 \leq \chi \leq 1$, the fidelity can be made above the classical limit by choosing proper finesse. The general trend indicates a trade-off between efficiency and fidelity. The shaded region is also where quantum cloning occurs if the photons in the echo pulses are treated as the clones of the input photon. An optical amplifier made of a fully inverted medium where the absorption is eliminated is an optimal quantum cloning machine [24, 25, 29]. According to equation (6) in ref [29], the numbers of clones in the same and orthogonal states of an input photon are, respectively, $\mu_V = 2\eta - 1$ and $\mu_H = \eta - 1$, where $\eta$ is the amplification gain corresponding to the echo efficiency for $\eta \geq 1$. These lead to an optimal fidelity $F_{opt} = (2\eta - 1)/(3\eta - 2)$ as plotted in Fig. 4 (dashed line). The quantum cloning in the AFC is not optimal due to the absorption from the ground state population that is required for echo generation. The point closest to the optimal curve is $F \approx 0.70$ at $\eta = 2.74$, $\chi = 1$, and $\Gamma/\gamma = 3$ while $F_{op} = 0.72$.

As expected, both efficiency and fidelity degrade with the background level. Fig. 5 shows the fidelity vs. efficiency for $b=0.1$. At this slightly elevated background, only a few curves fall into the quantum cloning regime (shaded area). Moreover, no curves are monotonic. The efficiency peaks roughly at $3<\Gamma/\gamma<4$, beyond which point the efficiency starts decreasing while the fidelity falls below the classical limit as well. The undecayed excited state still produces a curve closest to the optimal cloning.

According to equation (24), there is a general upper limit of the background level to ensure that $\eta \geq 1$. The necessary requirement is

$$\frac{(1+\chi)(1-b)}{b\Gamma/\gamma+(1-b)} \geq 1 \tag{26}$$

which can be simplified to

$$\frac{1-b}{b} \geq \frac{\Gamma}{\gamma\chi} \tag{27}$$

The minimum of the right hand side is 2 at $\frac{\Gamma}{\gamma}=2$ and $\chi=1$. This leads to a necessary condition $b \leq 1/3$ for a gain ($\eta \geq 1$) in the output compared with the input. The actual condition may be more stringent. For an AFC of square shaped tooth, the readout efficiency falls below 100% around $b=0.18$. Beyond this point, the memory readout efficiency is below unity and quantum cloning in the AFC loses its meaning.

## VI. Conclusion

A photon echo process in AFCs consisting of ground and excited state populations has been studied for the purpose of quantum memory and quantum cloning. A semi-classical theory has been adopted to study the absorption of the input and emission of photon echoes through a stimulated process under the condition of weak input interacting with the AFCs. Noise from the spontaneous emission due to atoms in the excited state has been included in the output. After the general forms of readout efficiency, SNR, and fidelity have been derived, the efficiency and fidelity were obtained for an input at the single photon level. The effects of the excited state population, as well as the spectral background of the AFC in the ground state have been analyzed for squared-tooth combs with various finesses. A quantum cloning region has been identified where the echo fidelity is larger than the classical limit while the readout efficiency is higher than unity. Quantum cloning is always possible in AFCs without background. However, Due to the absorption, the fidelity always stays below that of the optimal quantum cloning machine.

AFCs with undecayed excited states at finesse of about 3 have a fidelity closest to the optimum (0.70 vs. 0.72) at an efficiency of 270%. The uniform spectral background in the AFC is detrimental for both quantum memory and cloning. An upper limit of the background has been derived, beyond which the memory efficiency falls below unity and quantum cloning becomes impossible.

## ACKNOWLEDGMENTS

The authors wish to acknowledge the grant support from the National Science Foundation (PHY-1212360).

Captions:

(color online) Fig. 1 (a) A three-level atom with ground, excited, and metastable state labeled by $|g\rangle$, $|e\rangle$, and $|m\rangle$, respectively. (b) Spectral distribution of atomic population in ground and excited state corresponding to an AFC with square-shaped teeth.

(color online) Fig. 2 Readout efficiency vs. the background level $b$ and finesse $\Gamma/\gamma$ from a square-tooth AFC with an unoccupied excited state ($\chi=0$). The ten contour bands illustrate the efficiency increase, in the direction indicated by the arrow, from 0 to 100% with an increment of 10%.

(color online) Fig. 3 Fidelity vs. the background level and finesse from the same AFC as that in Fig.2 except for an undecayed excited state ($\chi=1$). The twenty contour bands illustrate the fidelity increase, in the direction indicated by the arrow, from 0.53 to to 0.73 with an increment of 0.1.

(color online) Fig. 4 Fidelity vs. readout efficiency from the AFC without background ($b=0$) at different excited decay fractions ($1-\chi$) and finesse. Ten dots on each solid curve represent the fidelity values for finesse ranging from 2 to 11 from left to right along the curve. The shaded area indicate the region where $\eta \geq 100$ and $F > 2/3$. The dashed line represents the optimal cloning.

(color online) Fig. 5 Fidelity vs. readout efficiency from the AFC with a background ($b=0.1$) at different excited decay fractions ($1-\chi$) and finesse. Ten dots on each solid curve represent the fidelity values for finesse ranging from 2 to 11 from top to bottom along the curve. The shaded area indicate the region where $\eta \geq 100$ and $F > 2/3$.

Fig.1

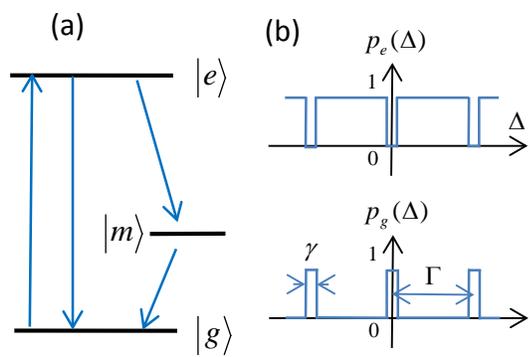

Fig.2

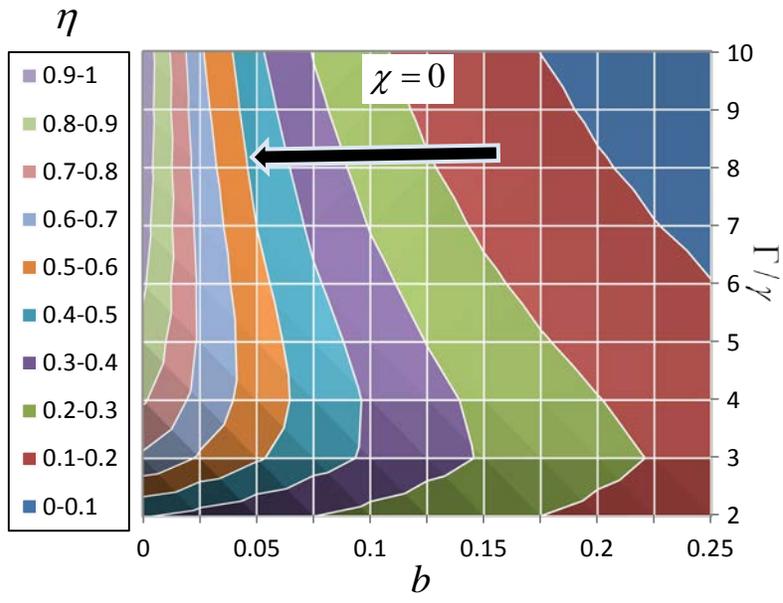

Fig. 3

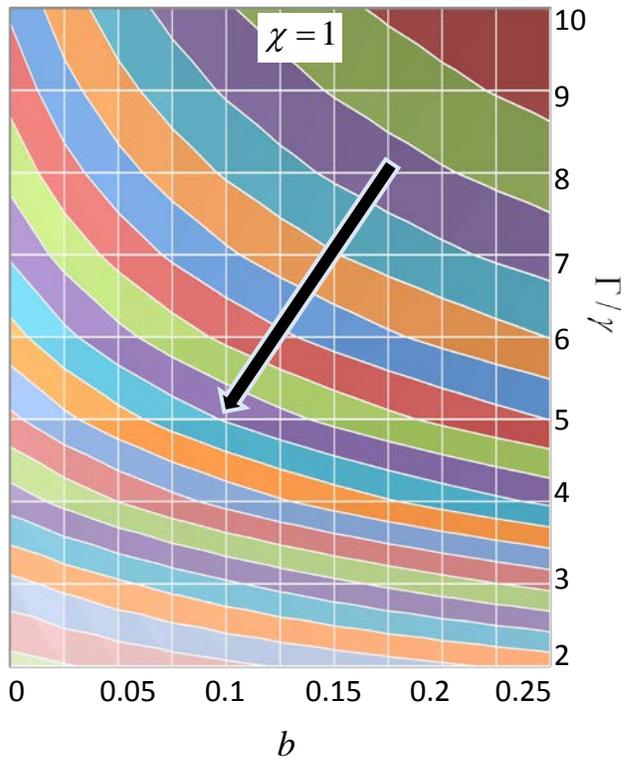

Fig. 4

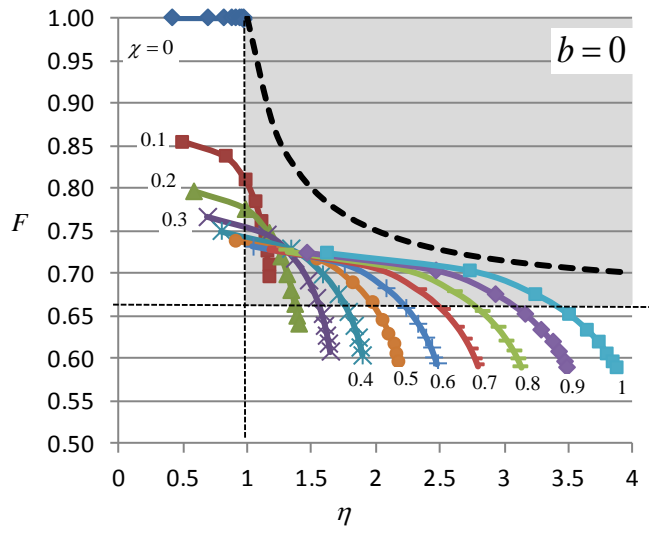

Fig. 5

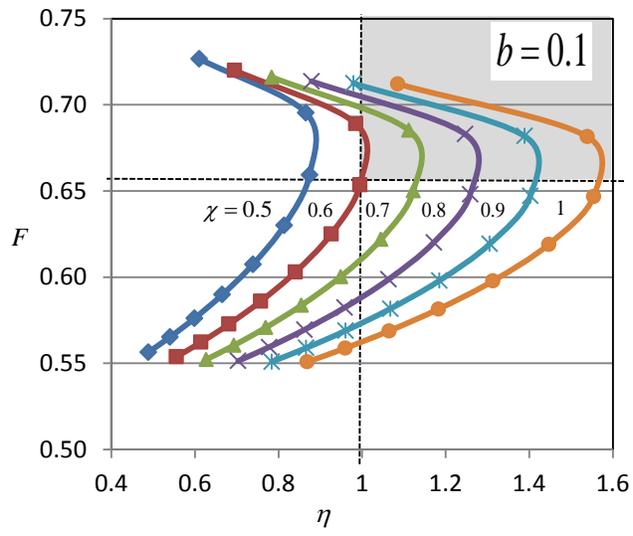